# Performance Analysis of Integrated Power-Line/Visible-Light Communication Systems with AF Relaying

Waled Gheth, Khaled M. Rabie, Bamidele Adebisi, Muhammad Ijaz and Georgina Harris
School of Engineering, Manchester Metropolitan University, Manchester UK
Emails:{w.gheth, k.rabie, b.adebisi, m.ijaz, g.harris}@mmu.ac.uk

*Abstract*—Reliable data transmissions and offering better mobility to the end user can be achieved by integrating different communication systems. In this paper, we investigate the performance of a cascaded indoor power line communication (PLC)/visible light communication (VLC) system with the presence of an amplify-and-forward (AF) relay. Using the pre-installed infrastructure of electricity wiring networks gives the advantage to use PLC as a backbone for VLCs. The performance of the proposed hybrid system is discussed in terms of the average capacity. A mathematical method is developed for this network to formulate the capacity by exploiting the statistical properties of both the PLC and VLC channels. The derived analytical expressions are validated by Monte Carlo simulations. The results showed that there is a considerable improvement in the performance of the hybrid system as the relay gain increases whereas it deteriorates with increasing the end-to-end distance. A comparison between the performance of a parallel hybrid/PLC and hybrid systems is also provided. It is found that the hybrid/PLC system outperforms the hybrid one. However, the user mobility offered by the latter system remains the main advantage over the former approach.

## I. INTRODUCTION

RECENTLY, integrating communication systems in order to ensure a reliable transmission, such as integrating power line communication (PLC) with visible light communication (VLC) and PLC with wireless communication systems, has been considered by many researchers [1], [2]. The implementation of such integration can provide several advantages in terms of capacity, security and data rate improvements [1]. The PLC system is considered one of the competitive technologies for broad-band communications in indoor applications as it exploits the pre-installed power lines network. In addition, the optical wireless communication (OWC) technology is predicted to play a crucial role in indoor wireless applications. The infra-red communication and VLC are the two main technologies of the indoor OWC. However, due to the widespread usage of the LEDs for their economic consumption of energy the VLC has attracted significant attention and it is considered as one of the most important green communication technologies [3], [4].

However, due to the nature of the channels of the cascaded system, different practical challenges should be considered while implementing such integration. For instance, electromagnetic interference (EMI) caused by the PLC link to the surroundings is one of these issues that needs to be regarded when designing the hybrid PLC/VLC system [5], [6]. Noise over such cascaded channels is another issue. For example, impulsive noise over PLC channel which has a random occurrence, short duration and high power spectral density (PSD) [7]–[10]. Furthermore, attenuation which is caused by the the frequency-selectivity of the PLC channel has an adverse effect on the transmission over such channels and can considerably degrade the achievable data rates. The attenuation is considerably high at very high frequencies and increases with the transmitter–receiver distance [11], [12]. In addition, interference caused by the overlap of adjacent LED light irradiation area in the VLC link is one of the technical problems. Moreover, instability of signal reception caused by the frequent switching in different LED coverage areas as the area coverage by each LED is small [13].

Motivated by adopting other communication technologies for PLC and VLC systems, many researchers have considered the implementation of relays for both systems to improve the performance of system [14]–[17]. The two common types of relaying systems are decode-and-forward (DF) and amplify-and-forward (AF). While the former decodes the received signal and re-encodes it before forwarding it to the destination, the received signal is amplified by the AF then re-transmitted to its destination [18]. Other types of relays investigated over PLC channels include incremental DF and selective DF [19]–[21].

The authors in [15] investigated the deployment of half duplex time division relaying protocols in indoor PLC systems and they concluded that the data rate and power saving were significantly improved depending on two main factors namely, the network size and relay position in the network. Although the results in [22] were very promising considering the position of the relay in the network, the authors reported that relaying in PLC systems is less efficient than it is for wireless networks. Authors in [14] reported that there is a significant performance improvement by using two-way relaying compared to the one-way relaying scenario. The authors in [23] discussed the implementation of multi-way relaying in parallel for indoor PLC systems and the achievable data rates were high compared to those achieved by direct transmission. The implementation of relays in hybrid communication systems was also studied in [1] where the source transmits signals through PLC and wireless networks simultaneously to a destination through an AF relay. It was reported that the use of hybrid systems outperforms the use of PLC or wireless systems individually.

The cascaded free space optics (FSO)-VLC communication system was discussed in [16] with a DF relay in between the FSO and VLC links. The proposed system was investigated in terms of capacity and the results showed that the hybrid system is more efficient in terms of data rate. The cooperation between light sources and how it can improve the performance of VLC systems was investigated in [24] where the authors reported that the implementation of DF relaying offers better performance than that of AF relaying.

To the best of our knowledge, no work has studied AF relays to connect PLC and VLC links. Therefore, this paper studies the implementation of AF relaying to connect a cascade PLC and VLC links. In the first phase, the PLC source sends the information signal to a relay through a PLC link. In the second phase, the AF relay amplifies and forwards the received signal to the destination node via the VLC channel. An analytical expression for the total capacity of the system is derived. Formulating the overall capacity of the hybrid PLC/VLC system offers the opportunity to examine the performance of such systems as well as to examine the effect of the different system parameters such as relay location and relay-to-destination distance. Monte Carlo simulations were used throughout the paper to verify the accuracy of the analytical results. For the sake of comparison, the performance of the hybrid system is compared with the performance of the hybrid/PLC, in which a linear combiner is used to maximize the overall capacity of the system.

The rest of the paper is organized as follows. The system model is described in Section II. The system performance analysis is presented in Section III. Discussions of the numerical results are produced in Section IV. Finally, Section V draws the conclusion of this paper.

## II. SYSTEM MODEL

In this section, the system model for the hybrid PLC/VLC system with the use of relay is presented and the overall capacity is formulated. The system model under consideration is shown in Fig. 1. PLC users can be connected to the network through PLC modems. We assume that the source node is on one floor and a VLC user is on another floor. The data sent by the source node through the PLC link is amplified and re-transmitted to the destination through the VLC link. The complex channel gains $h_P$ and $h_v$ represent the source-to-relay channel gain, (i.e., the PLC link) and the channel gain of the VLC link (i.e., the relay-to-destination channel), respectively. Both channels are assumed to be independent and identically distributed. The amplitude of the channel coefficients of the PLC channel is log-normal [25]. Although, the down-link transmission of the VLC link consists of line-of-sight (LOS) and non-LOS (NLOS) components, the LOS link is only considered in this study as it represents more than 90% of the total received signal [26]. The LED is located on the ceiling with a vertical distance to the user plane $L$ and Euclidean distance to the destination $d_k$. The VLC channel is subjected to a random distribution which is affected by the uniform distribution of the user's location [26]. Because of the nature of the network structure adopted here, it is logical to assume that there is no direct link between the source and destination nodes. It is worth mentioning, for simplicity and without loss of generality, that noise over the two links is assumed to be additive white Gaussian noise (AWGN).

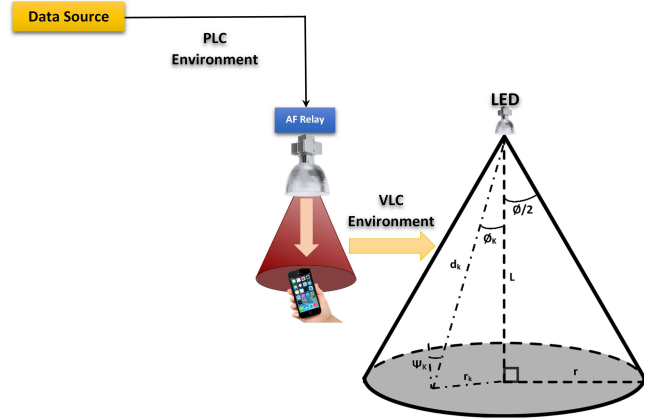

Fig. 1: System model for the hybrid PLC/VLC network for indoor communication applications.

The whole communication process as previously mentioned is divided into two phases. The signal received during the first segment at the relay is given by

$$y_r(t) = \sqrt{P_s} e^{-\alpha d_1} h_P s + n_r, \quad (1)$$

where $P_s$ represents the source transmit power, $d_1$ represents the source-to-relay distance of the PLC link length, $s$ is the information signal with $E[s] = 1$, $n_r$ is the noise at the relay which is assumed to be complex Gaussian with zero mean and variance $\sigma_r^2$, and $\alpha$ is the PLC channel attenuation factor and is given by $\alpha = a_0 + a_1 f^k$, where $a_0$ and $a_1$ are constants determined by measurements, $f$ is the system operating frequency and $k$ denotes the exponent of the attenuation factor.

The received signal at the VLC user can be described by the following formula

$$y_d(t) = h_v G \sqrt{P_r} \left( \sqrt{P_s} e^{-\alpha d_1} h_P s + n_r \right) + n_d, \quad (2)$$

where $G$ is the relay gain, $P_r$ is the relay transmit power and $n_d$ represents the destination noise with zero mean and variance $\sigma_d^2$.

Using (2), and grouping the information signal and noise signal terms, the signal-to-noise ratio (SNR) at destination can be written as

$$\gamma = \frac{G^2 P_r P_s e^{-2\alpha d_1} |h_P|^2 |h_v|^2}{G^2 P_r |h_v|^2 \sigma_r^2 + \sigma_d^2}. \quad (3)$$

The instantaneous capacity can be calculated as

$$C = \frac{1}{2} E\left[\log(1 + \gamma)\right], \quad (4)$$

where $\gamma$ is the SNR at the destination and the factor $\frac{1}{2}$ is due to the fact that communication between the transmitter and receiver is performed in two stages.

## III. PERFORMANCE ANALYSIS

In this section, we derive accurate analytical expressions of the average capacity for both the hybrid and PLC-only systems.

### A. Hybrid PLC/VLC System

To begin with, we rewrite (3) as

$$\gamma = \frac{P_s e^{-2\alpha d_1} |h_P|^2}{\sigma_r^2 + \frac{\sigma_d^2}{P_r G^2 |h_v|^2}}. \quad (5)$$

To make the analysis easier to follow, we use the following substitutions $A = P_s e^{-\alpha d_1} |h_P|^2$, $b = \sigma_r^2$ and $C = \frac{\sigma_d^2}{P_r G^2 |h_v|^2}$. Now we can write (5) as

$$\gamma = \frac{A}{b+C}. \quad (6)$$

Substituting (6) into (4), the hybrid system average capacity can be expressed as

$$C_H = \frac{1}{2} E\left[\log\left(1 + \frac{A}{b+C}\right)\right], \quad (7)$$

where the superscript $H$ denotes the hybrid system. According to [27, Eq (5)], for any non-negative random variable $u$ and $v$, we can calculate the average capacity as follows

$$E\left[\ln(1 + \frac{u}{v})\right] = \int_0^\infty \frac{1}{z}\left(1 - M_u(z)\right)\mathcal{M}_v(z)dz, \quad (8)$$

where $M_u(z)$ and $M_v(z)$ denote the moment generation functions (MGFs) of the random variables $u$ and $v$, prospectively. Using this definition, we can rewrite (7) as follows

$$E\left[\ln(1 + \frac{A}{b+C})\right] = \frac{1}{2}\int_0^\infty \frac{1}{z}\left(1 - M_A(z)\right) M_{C+b}(z)dz, \quad (9)$$

where $\mathcal{M}_A(z)$ denotes the MGF of the random variable $A$ given by

$$\mathcal{M}_A(z) = M_{|h_P|^2} P_s e^{-2\alpha d_1}, \quad (10)$$

and $\mathcal{M}_{C+b}(z)$ is the MGF of $C + b$.

As previously mention, the PLC channel is affected by log-normal fading and according to [28, Eq (2.54)] the MGF of the log-normal distribution is given by

$$\mathcal{M}_h(z) \simeq \frac{1}{\sqrt{\pi}} \sum_{n=1}^{N_p} H_{x_n} \exp\left(10^{\frac{-\sqrt{2}\sigma_h x_n + (-2\mu_h)}{10}} z\right), \quad (11)$$

where $H_{x_n}$ and $x_n$ represent the weight factors and zeros of the $N_p$ order Hermite polynomial, respectively. By using this for $|h_P|^2$ with parameters $(2\mu_{h_P}, 4\sigma_{h_P}^2)$, the MGF of $A$ can be expressed as

$$\mathcal{M}_A(z) \simeq \frac{1}{\sqrt{\pi}} \sum_{n=1}^{N_p} H_{x_n}$$
$$\exp\left(10^{\frac{-\sqrt{2}2\sigma_{h_P} x_n + (-2\mu_{h_P})}{10}} P_s e^{-2\alpha d_1} z\right). \quad (12)$$

On the other hand, considering the random distribution mentioned above for the VLC link and the Lambertian radiation pattern for LED light emission, the VLC channel gain $h_v$ can be written as [16]

$$h_v = \frac{m_k + 1}{2\pi d_k^2} A_d \cos^{m_k}(\phi)\cos(\Psi_K) U(\Psi_K) g(\Psi_K) R_p, \quad (13)$$

where $A_d$ is the detection area of the detector, $d_k = \sqrt{r_k^2 + L^2}$, $U(\Psi_K)$ represents the optical filter gain, $g(\Psi_K)$ denotes the optical concentration gain, the responsivity of the photo-detector is represented by $R_p$, $\phi$ is the total angle of the LED, $\cos^{m_k}(\phi) = \cos(\Psi_K) = \frac{L}{\sqrt{r_k^2 + L^2}}$, and $m_k$ is the order of the Lambertian radiation pattern which is given by

$$m_k = \frac{-1}{\log(\cos(\phi/2))}, \quad (14)$$

where $\phi/2$ denotes the semi-angle of the LED.

To simplify our analysis, let us assume that $Q = \frac{1}{2\pi} A_d U(\Psi_K) g(\Psi_K) R_p$. In light of this, (13) can be rewritten as

$$h_v = \frac{Q(m_k + 1) L^{m+1}}{(r_k^2 + L^2)^{\frac{m+3}{2}}}. \quad (15)$$

It is assumed that the location of the users is uniformly distributed with probability density function (PDF) given as

$$f_{r_k}(r) = \frac{2r}{r^2}. \quad (16)$$

The PDF of the un-ordered channel gain of the VLC link can be driven using the change-of-variable method used in [16] as follows

$$f_{h_k}(h) = |\frac{\partial}{\partial h} u^{-1}(h)| f_{h_k}(u^{-1}(h)). \quad (17)$$

In this respect, we can now express the PDF of the VLC channel gain as

$$f_{h_k} = \frac{2Q^{\frac{2}{2+m}} \left((m_k + 1) L^{m_k + 1}\right)^{\frac{2}{m+3}} h^{-\frac{2}{m_k+3}-1}}{(m_k + 3)r^2}, \quad (18)$$

where $r$ is the maximum cell radius of the LOS.

The PDF of $h_k^2$ can consequently be obtained as follows

$$f_{h_k^2} = \frac{-Q^{\frac{2}{2+m}} \left((m_k + 1) L^{m_k + 1}\right)^{\frac{2}{m+3}} h^{-\frac{m_k+5}{m_k+3}}}{(m_k + 3)r^2}. \quad (19)$$

In order to obtain the MGF for $C + b$, we have to find the PDF of $\frac{1}{|h_v|^2}$ as $C = \frac{\sigma_d^2}{P_r G^2 |h_v|^2}$. Using the change-of-variable method the PDF of $\frac{1}{|h_v|^2}$ can be written as

$$\mathcal{M}_{\frac{1}{h_v^2}}(z) = \frac{\left(Q\left(m_k+1\right)L^{m_k+1}\right)^{\frac{2}{m+3}}}{(m_k+3)r^2}\left(z^{-\frac{1}{m_k+3}}\Gamma\left(\frac{1}{m_k+3},\frac{L^4 z}{Q^2(1+m_k)^2}\right)\right.$$
$$\left.-z^{-\frac{1}{m_k+3}}\Gamma\left(\frac{1}{m_k+3},\frac{L^{-2(m_k+1)}(L^2+r^2)^{m_k+3}z}{Q^2(1+m_k)^2}\right)\right). \quad (22)$$

$$\mathcal{M}_{C+b}(z) = \frac{(\frac{\sigma_d^2}{P_r G^2}z)^{-\frac{1}{m_k+3}}}{(m_k+3)r^2}\left(Q\left(m_k+1\right)L^{m_k+1}\right)^{\frac{2}{m+3}}\left(\Gamma\left(\frac{1}{m_k+3},\frac{L^4\sigma_d^2 z}{Q^2(1+m_k)^2 P_r G^2}\right)\right.$$
$$\left.-\Gamma\left(\frac{1}{m_k+3},\frac{L^{-2(m_k+1)}(L^2+r^2)^{m_k+3}\sigma_d^2 z}{Q^2(1+m_k)^2 P_r G^2}\right)\right). \quad (23)$$

$$f_{\frac{1}{h_v^2}} = \frac{\left(Q(m_k+1)L^{m_k+1}\right)^{\frac{2}{m+3}} h^{-\frac{2}{m_k+3}-1}}{(m_k+3)r^2}. \quad (20)$$

Now, by using (20) we can obtain the MGF of $\frac{1}{|h_v|^2}$ as follows

$$\mathcal{M}_{\frac{1}{h_v^2}}(z) = \int_{t_{min}}^{t_{max}} f_{\frac{1}{h_v^2}}(z)\exp(-zh)\,dz, \quad (21)$$

where $t_{min} = \frac{\left(Q(m_k+1)L^{m_k+1}\right)^2}{(r^2+L^2)^{m_k+3}}$ and $t_{max} = \frac{\left(Q(m_k+1)L^{m_k+1}\right)^2}{L^{2(m_k+3)}}$. Using (20) in (21), we obtain (22), shown at the top of the page.

Finally, the $\mathcal{M}_{C+b}(z)$ can be given as in (23), also shown at the top of this page.

### B. PLC System

In this subsection, we consider the scenario when the transmitter and the receiver are only connected through a PLC link with no relay in between. Therefore, the signal at the destination can be given by

$$y_d(t) = \sqrt{P_s}e^{-\alpha d}hs + n_d, \quad (24)$$

where $d$ is the PLC link length which represents the total distance of the end-to-end nodes and $n_d$ is the destination noise which is assumed to be complex Gaussian with zero mean and variance $\sigma_d^2$,

The SNR at the destination can be written as follows

$$\gamma = \frac{P_s e^{-2\alpha d}|h|^2}{\sigma_d^2}. \quad (25)$$

Mathematically, we can calculate the average capacity of this system as

$$C_P = \int_0^\infty \log_2(1+\gamma) f_\gamma(\gamma) d\gamma, \quad (26)$$

where $f_\gamma(\gamma)$ is the PDF of $\gamma$ given by

$$f_\gamma(\gamma) = \frac{\zeta}{z\sqrt{8\pi}\sigma}\exp\left(-\frac{(\zeta\ln(\gamma)-(2\mu+\zeta\ln(a_1)))^2}{8\sigma^2}\right), \quad (27)$$

where $\zeta$ is the scaling constant and it is equal to $10/\ln(10)$ and $a_1 = \frac{P_s e^{-2\alpha d}}{\sigma_{d2}^2}$. Hermite–Gauss quadrature is used in order to get an accurate approximation for (26). Assume that

$$x = \frac{\zeta\ln(\gamma)-2\mu+\zeta\ln(a_m)}{8\sigma^2}. \quad (28)$$

Now, we can rewrite (26) as

$$C_P = \int_\infty^\infty \frac{1}{\pi}h(x)\exp[-x^2]dx. \quad (29)$$

By using Hermite–Gauss quadrature (29) can be expressed as

$$C_P = \sum_{n=1}^{N_p} \frac{1}{\sqrt{\pi}} H_{x_n} h(x_n), \quad (30)$$

where

$$h(x_n) = \log_2\left(1+\exp\left(\frac{\sqrt{8}\sigma x_n + 2\mu + \zeta ln\phi/2(a_1)}{\zeta}\right)\right). \quad (31)$$

It is worth noting that similar analysis is followed in [29].

## IV. NUMERICAL RESULTS

This section presents some numerical results of the derived expression along with Monte Carlo simulations. The system parameters considered in our evaluations are, unless indicated otherwise, as follows: operating frequency of the system $f = 500$kHz, $k = 0.7$, $a_0 = 2.03 \times 10^{-3}$, $a_1 = 3.75 \times 10^{-7}$, $d_1 = 30$m, the input power $P_s = 1$W, $P_r = 1$W, relay gain $G = 1$, input SNR is 10dB, $A_d = 0.1$m, $U(\Psi_K) = g(\Psi_K) = 7$dB, $R_p = 0.4$A/W, $r_e = 3.6$m, $L = 2.15$m and $\phi/2 = 60°$.

To demonstrate the impact of the source-to-relay distance on the system performance, the average capacity is plotted in Fig. 2 versus the input power for different values of PLC link lengths. It is obvious that there is a perfect agreement between the simulated and the analytical results, obtained from (9). It is clear that the capacity improves as the input power increases. On the other hand, a performance degradation can be clearly seen when the distance between the source and relay increases and input power decreases.

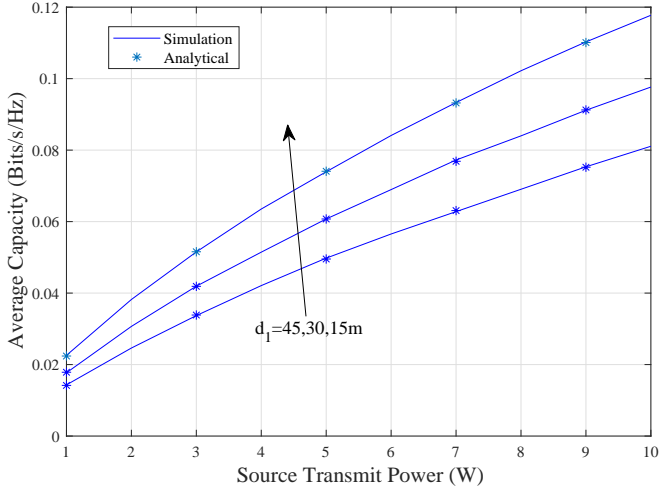

Fig. 2: Average capacity versus input power for different values of source-relay distance.

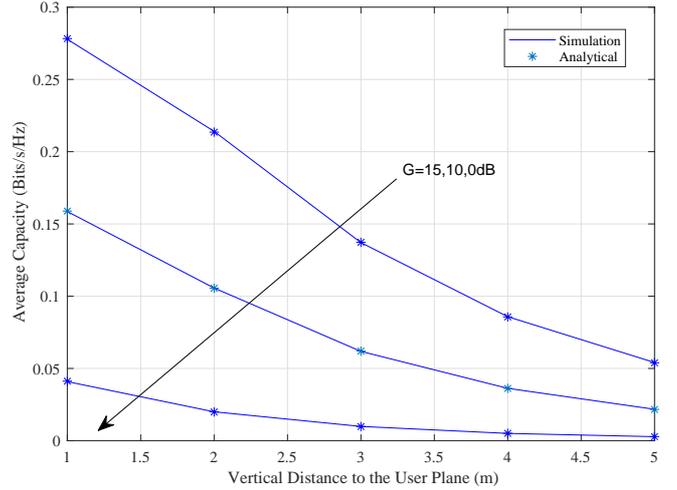

Fig. 4: Average capacity with respect to the vertical distance to the user plane for different values of the relay gain.

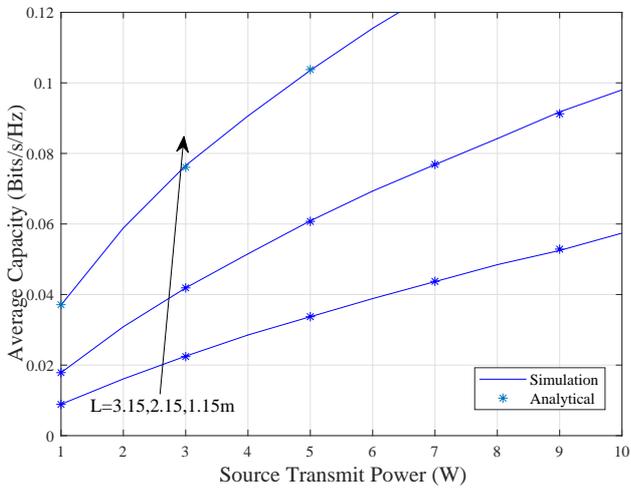

Fig. 3: Average capacity as a function of input power for different values of the vertical distance to the user plane.

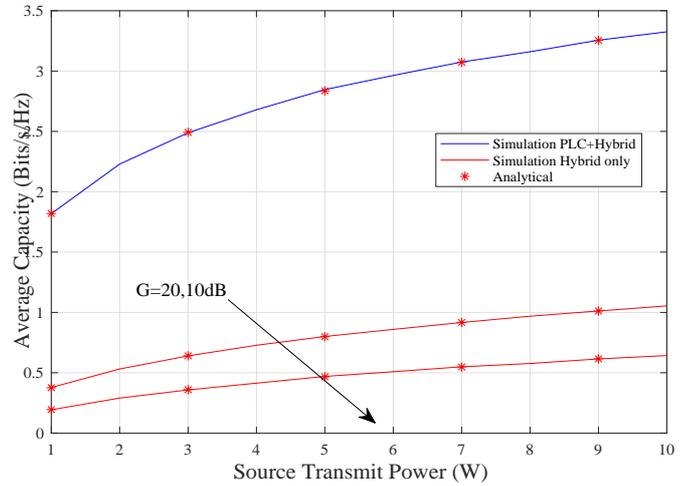

Fig. 5: Comparison between the hybrid system and PLC + hybrid system.

In Fig. 3, the total capacity is again plotted against the input power with different values of the vertical distance between the relay and the user plane. As anticipated, we can see that the average capacity is degraded as the vertical distance to the user plane increases. For instance, the average capacity is only $0.04$Bits/s/Hz when $L = 3.15$m and $P_s = 7$W, and the values of the average capacity are $0.08$Bits/s/Hz when $L = 2.15$m and $0.12$Bits/s/Hz when the vertical distance is $1.15$m for the same value of the input power.

Fig. 4 illustrates what happens to the average capacity when the gain of the relay increases. It is noticeable that the performance of the system has a significant improvement when the relay gain increases and this justifies the main purpose of the relay used in the system. As seen in the figure, the average capacity is very low when the relay gain is 0dB compared to its values for higher relay gains. It is obvious that the system becomes more efficient when the relay gain increases and the vertical distance between the relay and the user plane decreases.

In Fig. 5, we assumed that the destination has the capability to receive signals from both the hybrid and the PLC systems. For the sake of comparison, the average capacity of the hybrid and the hybrid/PLC systems is compared in Fig. 5. It is noticeable that the performance of the parallel hybrid/PLC is by far better than that of the hybrid system. Although we increased the relay gain in the hybrid system, yet the hybrid/PLC outperforms the hybrid system for all the study cases.

## V. CONCLUSIONS

In this paper, we discussed the implementation of AF relaying in hybrid PLC/VLC communication systems for indoor applications. The system performance was analyzed in terms of the average capacity for various system setups. The derived analytical expressions were validated by Monte Carlo simulations. The impact of several system parameters was investigated. The results showed that the relay gain has a positive effect on the performance of the proposed hybrid system. The capacity is also affected by the vertical distance between the LED and the user plane. It was found that the source-to-relay distance does not have a considerable effect on the performance if the increase of the distance is relatively small. Although, the performance of the hybrid/PLC system outperforms that of the hybrid system, the use of the hybrid system is more attractive in indoor applications since it provides better mobility to end users.

## VI. Acknowledgment

This research has been jointly funded by OSL Rail Ltd, and the Faculty of Science and Engineering, Manchester Metropolitan University, UK.